\documentclass[twocolumn,showpacs,amsmath,amssymb,prl]{revtex4}%

\usepackage{graphicx}
\usepackage{psfrag}
\usepackage{dcolumn}
\usepackage{bm}

\begin{document}

\preprint{PRL/Version 4}

\newcommand{\forex}{\emph{forex} }

\title{Foreign exchange market fluctuations as random walk
in demarcated complex plane}

\author{Johnrob Bantang}%
\author{May Lim}%
\author{Patricia Arielle Castro}%
\author{Christopher Monterola}%
\author{Caesar Saloma}%
 \email{csaloma@nip.upd.edu.ph}
\affiliation{%
National Institute of Physics, University of the Philippines\\
Diliman, Quezon City
}%

\date{\today}%

\begin{abstract}
We show that time-dependent fluctuations $\{\Delta x\}$ in foreign
exchange rates are accurately described by a 
random walk in a complex plane that is demarcated into the gain
($+$) and loss ($-$) sectors. $\Delta x$ is the outcome of $N$
random steps from the origin and $|\Delta x|$ is the square
of the Euclidean distance of the final $N$-th step position. Sign
of $\Delta x(t)$ is set by the $N$-th step location in the plane.
The model explains not only the exponential statistics of the
probability density of $\{\Delta x\}$ for G7 markets but also
its observed asymmetry, and power-law dependent broadening
with increasing time delay.
\end{abstract}


\maketitle

Easier data access via the Internet and the widespread
availability of powerful computers, have enabled many researchers
not just in business and economics \cite{ref1} but
also in physics, applied mathematics, and engineering \cite{ref5,
ref6, ref7, ref8}, to
investigate more deeply the dynamics of foreign exchange markets.
Their efforts have led to new insights on the characteristics of
\forex rate fluctuations including the general behavior of their
corresponding probability density function (\emph{pdf}) \cite{ref5, ref7}.

Arguably, \forex markets have a more immediate and direct impact
on citizens than stock markets do. This is especially true in
developing countries where only a minority holds stocks directly
while a majority sells or purchases products and commodities on a
daily basis. Economies that rely strongly on remittances of
overseas contract workers (e.g. Philippines) or tourism (e.g.
Thailand), are also quite sensitive to \forex rate instabilities.
Hence, knowledge of the vagaries of the \forex market, is
important not only to policymakers and macroeconomic
managers in government and banking sectors but also to individual
businessmen and consumers.

The exchange rate of a currency relative to another
(usually the US dollar) is represented by a time-series of
quotations $\{x(t)\}$, where time $t = m\Delta t$, $\Delta t =$
time delay, and index $m = 1, 2, \ldots, M$. Normally, the
power spectrum of $\{x(t)\}$ consists of one or two dominant
low-frequency components and a continuous high-frequency band. The
low-frequency components reveal the longterm behavior of
$\{x(t)\}$ and therefore the soundness of a national economy from
the perspective of economic fundamentals. The
high-frequency components are associated with fluctuations in
$\{x(t)\}$ that arise from complex interactions between market
players over short time-scales ($\Delta t < 2$ days) \cite{ref5,
ref7}.

\begin{figure}
\centering
  \includegraphics[width=\columnwidth]{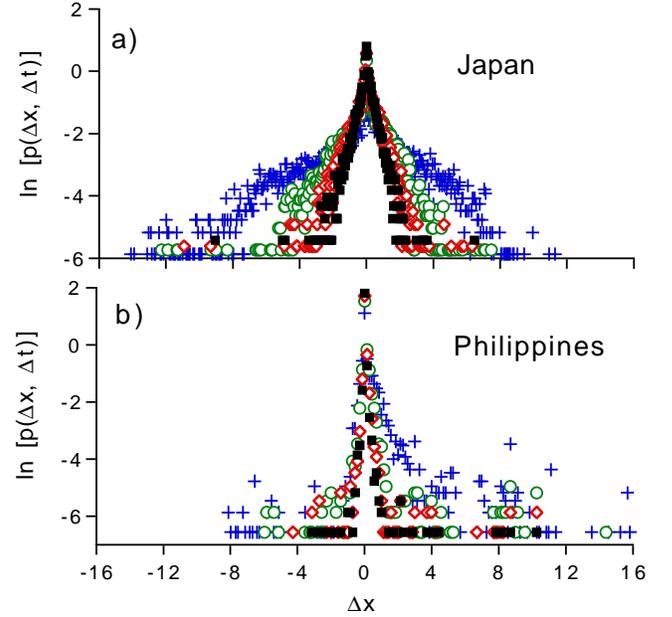}\\
  \caption{
Asymmetric $p(\Delta x, \Delta t)$'s: a) Japan, and b)
Philippines for $\Delta t$ (day) $= 1$ (filled squares), $2$
(diamonds), $5$ (circles), and $20$ (cross-hairs). For any $\Delta
t$, $p(\Delta x, \Delta t)$ is normalized i.e., $\sum p(\Delta x,
\Delta t)\delta x = 1$, where histogram bin-width $\delta x =
(2/201)[{\rm max}(\Delta x_{\rm max}, |-\Delta x_{\rm min}|)]$.
Summation is for all $\Delta x$'s.
  }\label{fig:1}
\end{figure}

The statistics of the (percentage) rate fluctuations $\{\Delta
x(t)\} = \{\Delta x\}$ in G7 markets is described by an
exponentially-decaying \emph{pdf} $p(\Delta x,
\Delta t)$ where: $\Delta x(t) = [x(t + \Delta t) -
x(t)][100/x(t)]$. This statistical behavior is not easily seen
from the power spectrum of $\{x(t)\}$. Figure~\ref{fig:1} shows
the $p(\Delta x, \Delta t)$'s for the Japanese (\ref{fig:1}a) and
Philippine (\ref{fig:1}b) markets. For the G7 \cite{ref14} and
Philippine \cite{ref15} markets, our core data sets for $\{x(t)\}$
comprise of daily closing rates from January 1971 to December 1993
($M = 5000$ trading days, $\Delta t = 1$ day). The Philippine
market was under exchange control in the 70's and its $p(\Delta x,
\Delta t)$ has a large number of points around $\Delta x = 0$.

Most $p(\Delta x, \Delta t)$'s exhibit a degree of asymmetry about
$\Delta x = 0$. An asymmetric $p(\Delta x, \Delta t)$ is described
by a pair of exponents $\alpha_{l}$ and $\alpha_{r}$ for $\Delta x
< 0$ (left) and $\Delta x > 0$ (right), respectively. It is
displayed by currencies that have either appreciated (e.g. Japan)
or depreciated (e.g. Canada, Philippines) against the USD during
the sampling period.

Table~\ref{tab:1} lists the best-fit values (by least-squares
method) of $\alpha_{l}$, $\alpha_{r}$ and the left and right
intercepts $\beta_{l}$ and $\beta_{r}$ of $p(\Delta x, \Delta t =
1\;{\rm day})$ with the $\Delta x = 0$ line for the G7 markets.
The exponential behavior of $p(\Delta x, \Delta t)$ persists even
for $\Delta t$'s $> 2$ days. We also found that $p(\Delta x,
\Delta t)$ broadens with $\Delta t$ in a power-law trend.
\begin{table*}
  \centering
  \caption{
Parameter values of exponential fit: $\beta \exp(\alpha \Delta
x)$, to $p(\Delta x, 1$ day$)$ for G7 and Philippine markets. For
a symmetric $p(\Delta x, \Delta t)$: $\alpha_{l} = |\alpha_{r}|$,
and $\ln(\beta_{l}) = \ln(|\beta_{r}|)$.
  }\label{tab:1}
\begin{ruledtabular}
\begin{tabular}{ccccc}
Country & $\alpha_{l}$ & $\alpha_{r}$ & $\ln(\beta_{l})$ &
$\ln(\beta_{r})$\\
\hline Canada & $6.143\pm0.197$ & $-5.807\pm0.150$ &
$1.090\pm0.071$ & $1.037\pm0.058$
\\
France & $2.283\pm0.082$ & $-2.469\pm0.096$ & $0.056\pm0.065$ &
$0.181\pm0.069$
\\
Germany & $2.166\pm0.076$ & $-2.327\pm0.068$ & $0.058\pm0.062$ &
$0.125\pm0.050$
\\
Italy & $2.316\pm0.113$ & $-2.44\pm0.102$ & $-0.012\pm0.090$ &
$0.152\pm0.074$
\\
Japan & $2.371\pm0.115$ & $-2.635\pm0.084$ & $0.073\pm0.092$ &
$0.187\pm0.062$
\\
UK & $2.171\pm0.091$ & $-2.244\pm0.086$ & $0.004\pm0.074$ &
$0.063\pm0.066$
\\
Philippines & $4.711\pm0.904$ & $-2.892\pm1.063$ &
$-1.745\pm0.564$ & $-1.464\pm0.964$
\\
\end{tabular}
\end{ruledtabular}
\end{table*}

The behavior of $\{x(t)\}$ has been described previously as a
Markov process with multiplicative noise and the emergence of a
highly-symmetric $p(\Delta x, \Delta t)$ such as that of the
German market \cite{ref5, ref7}, is explained by the
Fokker-Planck equation. However, these studies did
not explain the emergence of asymmetric $p(\Delta x, \Delta t)$'s
and the power-law behavior of their broadening with $\Delta t$.

\begin{figure}
\centering
\psfrag{Q}{$\Theta$}
\psfrag{Q=0}{$\Theta=0$}
  \includegraphics[width=\columnwidth]{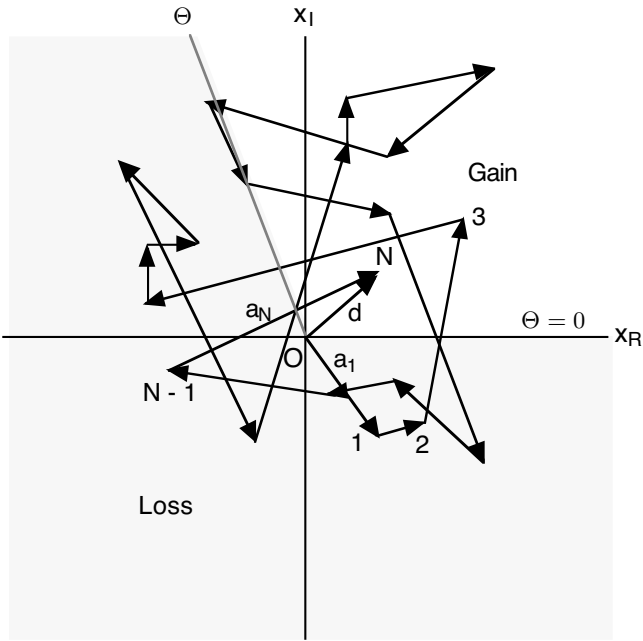}\\
  \caption{
Random walk in complex plane demarcated by angle $\Theta$, into
gain ($+$) and loss ($-$, shaded) regions. Each $\Delta x$ is the
outcome of N random steps from origin O where: $d =
\left(x_{R}^{2} + x_{I}^{2}\right)^{1/2}$, and $|\Delta x| =
d^{2}$. Sign of $\Delta x$ is given by location of endpoint
N($x_{R}$, $x_{I}$) in the plane. The n-th phasor is the
fluctuation $|\Delta x_{n}| = |a_{n}|^{2}$, of the $n$-th dealer
where $n = 1, 2, …, N$.
  }\label{fig:2}
\end{figure}

Here, we show that the statistics of $\{\Delta x\}$ is
accurately described by a random walk on a demarcated complex
plane. Our model accounts for possible asymmetry of $p(\Delta x,
\Delta t)$, its broadening with increasing $\Delta t$, and the
power-law behavior of this broadening for G7 markets. Each $\Delta x$ is the
outcome of $N$ random steps that start from the plane origin O
(see Fig~\ref{fig:2}). Its magnitude is $|\Delta x| = d^{2}$,
where $d$ is the Euclidean distance of the $N$-th step endpoint
N($x_{R}, x_{I}$) from origin O, and $x_{R}$ and $x_{I}$ are the
real and imaginary coordinates. The sign of $\Delta x$ depends on
the location of N($x_{R}, x_{I}$) in the plane which is demarcated
by the polar angle $\Theta$, into two regions representing the
gain ($+$) and loss ($-$) regimes.

\emph{Random walk in complex plane.} The
location of N($x_{R}, x_{I}$) is given by the sum of $N$
elementary phasors $\{a_{n}\}$ \cite{ref16, ref17}:
%
$    {\rm N}(x_R,x_I) = 
\sum_{n=1}^{N}\; a_{n}(x_R, x_I)
= \sum_{n=1}^{N}\; |a_{n}|\exp(j\; \theta_n)
$
%
where $j = (-1)^{1/2}$. The amplitude $|a_{n}|$ and phase
$\theta_{n}$ of the $n$-th phasor are statistically independent of
each other and of the amplitudes and phases of all other ($N - 1$)
$a_{n}$'s. Possible $|a_{n}|$ values are uniformly distributed
within a predetermined (non-negative) range.
Phases $\theta_{n}$ are also uniformly distributed:
$0 \leq \theta_{n} \leq 2\pi$.

Consistent with the central-limit theorem, the location of
N($x_{R}, x_{I}$) is governed by a Gaussian \emph{pdf}
($N \rightarrow \infty$) \cite{ref16}: $p_{N}(x_{R}, x_{I}) =
(2\pi \sigma^{2})^{-1} \;\exp [-(x_{R}^{2} +
x_{I}^{2})/2\sigma^{2}]$, where $\sigma^{2}$ is the variance.
Hence, the $d^{2}$-values obey a negative-exponential statistics:
 $p_{d}(d^{2}) = (2\sigma^{2})^{-1}
\exp (-d^{2}/2\sigma^{2})$. The $q$-th moment $\langle (d^{2})^{q}
\rangle$ of $d^{2}$ is given by: $\langle (d^{2})^{q} \rangle =
q!(2\sigma^{2})q = q! \langle d^{2} \rangle^{q}$, where the mean
value $\langle d^{2} \rangle$ is $2\sigma^{2}$. Phase $\theta$ of
N($x_{R}, x_{I}$) obeys a uniform statistics: $p_{\theta}(\theta)
= (2\pi)^{-1}$ for $-\pi \leq \theta \leq \pi$;
$p_{\theta}(\theta) = 0$, otherwise.

The joint (second-order) \emph{pdf} of
$d_{1}^{2}$ and $d_{2}^{2}$ at two different
time instants is \cite{ref16}: $p_{d}(d_{1}^{2}, d_{2}^{2}) =$
$\left[\langle d^{2} \rangle^{2} (1 - |\mu|^{2}) \right]^{-1} \exp
\left\{- \left( d_{1}^{2} + d_{2}^{2} \right) / \left[ \langle
d^{2} \rangle (1 - |\mu|^{2}) \right] \right\}$
$J_{0}\left\{2d_{1}d_{2}|\mu| / \left[\langle d^{2} \rangle (1 -
|\mu|^{2}) \right]\right\}$, where: $\langle d_{1}^{2} \rangle =
\langle d_{2}^{2} \rangle = \langle d^{2} \rangle$, $J_{0}(x)$ is
the zero-order Bessel function of the first kind, and $|\mu|$ is
the modulus of a complex factor that measures the degree of
correlation between events at two different time-instants. If
$|\mu| = 0$ (no correlation), $p_{d}(d_{1}^{2}, d_{2}^{2}) =
p_{d}(d_{1}^{2})\; p_{d}(d_{2}^{2})$. On the other hand,
$p_d(d_{1}^{2}, d_{2}^{2}) = p_{d}(d_{1}^{2})\; \delta(d_{1}^{2} -
d_{2}^{2})$, as $|\mu| \rightarrow 1$, where $\delta(x)$ is the
Dirac delta function.

As it is, the statistics of a random walk in a complex plane is
insufficient to describe the characteristics of the $p(\Delta x,
\Delta t)$'s of \forex markets because the possible $\Delta x$'s
are from $-\infty$ to $\infty$, while $d^{2} \geq 0$.

\emph{Demarcated complex plane.}
We solve the problem by demarcating the complex plane into two
sectors representing the gain ($+\Delta x$) and loss ($-\Delta x$)
regimes, where we identify that: $|\Delta x| = d^{2}$. The gain
area is set by the polar angle $\Theta$ (counterclockwise
rotation) and $\Delta x$ is positive (negative) if N($x_{R},
x_{I}$) is on the gain (loss) sector. The gain (loss) area is
zero if $\Theta = 0$ ($\Theta = 2\pi$).

\begin{figure}
\centering
  \includegraphics[width=\columnwidth]{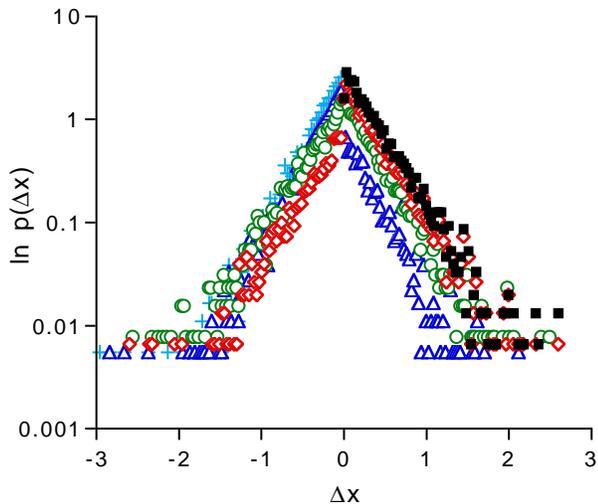}\\
  \caption{
\emph{Random walk in demarcated complex plane.} Histograms of
$\Delta x$'s for $\Theta = 0$ (squares), $\pi/2$ (diamonds), $\pi$
(circles), $3\pi/2$ (triangles) and $2\pi$ (crosses) where N $=
50$, $0 \leq |a_{n}| \leq 0.015$ (double precision).
  }\label{fig:3}
\end{figure}

Figure~\ref{fig:3} presents exponentially-decaying histograms of
$\Delta x$'s generated at $\Theta = 0$, $\pi/2$, $\pi$, $3\pi/2$, and $2\pi$ (360 deg). The
histograms are asymmetric about $\Delta x = 0$ at all angles
except at $\Theta = \pi$ where the plane is equally divided
between the gain and loss sectors. The corresponding $\Theta$'s
for the G7 markets in Table~\ref{tab:1} are not easily determined
since $\Theta$ is not clearly related with $\alpha_{l}$,
$\alpha_{r}$, $\beta_{l}$, and $\beta_{r}$. However, our model
reveals a unique relation between $R = A_{l}/A_{r}$ and $\Theta$,
where $A_{l}$ ($A_{r}$) is the area under the best-fit exponential
for $\Delta x < 0$ ($\Delta x > 0$) of the histogram.

\begin{figure}
\centering
  \includegraphics[width=\columnwidth]{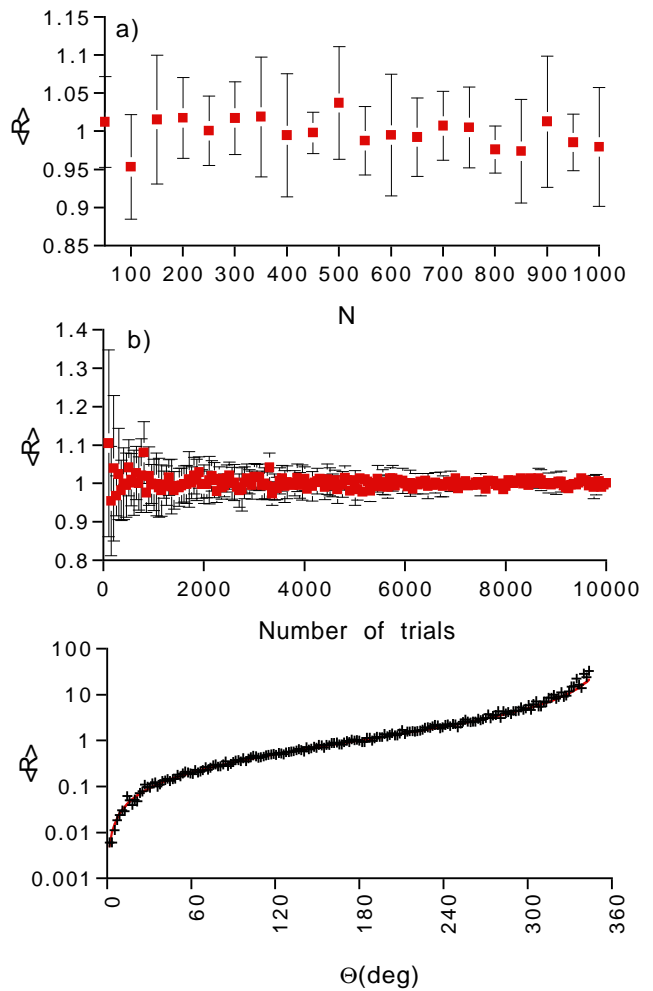}\\
  \caption{
$\langle R \rangle$ as a function of ($\Theta = \pi$): a) $N$, b)
Number of trials per datapoint ($N = 50$) ($\langle$s.d.$\rangle$ $=
1.05$.), and c) $\Theta$ for $N
= 50$. $10$ trials for each datapoint. 
  }\label{fig:4}
\end{figure}

Figures~\ref{fig:4}a-b plot the average $\langle R \rangle$ as a
function $N$ and number of trials,
respectively ($\Theta = \pi$). $\langle R \rangle$ is insensitive
to $N$ but for a fixed $N$, the standard deviation of $\langle R
\rangle$ decreases quickly (power-law decay) with the number of
trials . Figure~\ref{fig:4}c shows that the dependence of $\langle
R \rangle$ with $\Theta$ is well-described by: $\langle R \rangle
= \Theta / (360\; {\rm deg} - \Theta)$.

The $\langle R \rangle$'s are calculated from $p(\Delta x, \Delta
t)$'s of the G7 markets and then used to determine their
corresponding $\Theta$'s via the $R(\Theta)$-curve. The following
$\Theta$'s (deg) were obtained ($\Delta t = 1$ day): Canada
($179.7$), France ($180.4$), Germany ($182.3$), Italy ($177.8$),
Japan ($177.6$), and UK ($174.9$). For the G7 markets, differences
in the size of the gain and loss sectors are small ($< 4\%$). On
the other hand, the $p(\Delta x, 1\; {\rm day})$ for the
Philippine market is highly asymmetric with a significantly small
gain region ($\Theta = 108.1$ deg). Against the USD, the
Philippine peso depreciated by $246\%$ from December 1983 (lifting of exchange control) to
December 1993.

\emph{Broadening of $p(\Delta x, \Delta t)$ with time delay
$\Delta t$.} For the G7 markets, $p(\Delta x, \Delta t)$
broadens with increasing $\Delta t$, while preserving its
original negative exponential statistics within: $1 \leq
\Delta t\; ({\rm day}) \leq 20$. Previous studies on broadening
were confined to $\Delta t < 12$ hours \cite{ref7}.

\begin{figure}
\centering
  \includegraphics[width=\columnwidth]{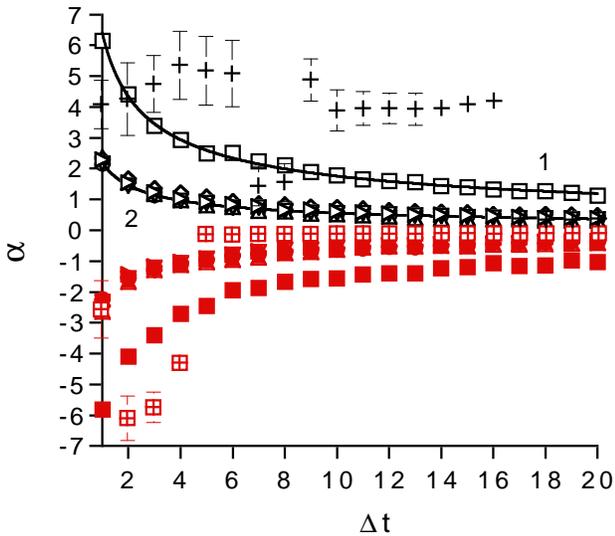}\\
  \caption{
Dependence of $\alpha_{l}$ (positive) and $\alpha_{r}$
(negative) with $\Delta t$ for Canada (squares for $\alpha_{l}$,
solid squares for $\alpha_{r}$), France (right triangles), Germany
(inverted triangles), Italy (diamonds), Japan (triangles), UK
(circles) and Philippines (crosses for $|\alpha_{l}|$, crossed
squares for $\alpha_{r}$). Reference curves:
$\alpha_{l} = 6.4 \Delta t-0.561$ (Curve 1) and $\alpha_{l} = 2.21
\Delta t - 0.593$ (Curve 2). $p(\Delta x_{Q}, Q)$
contains $5000$ datapoints and each point is the average of $10$
trials. 
  }\label{fig:5}
\end{figure}

Figure~\ref{fig:5} plots the dependence of $\alpha_{l}$ and
$\alpha_{r}$ with $\Delta t$ for the G7 markets where a power-law
behavior is observed. Except for Canada, the dependence of
$\alpha_{l}$'s (and $|\alpha_{r}|$'s) is remarkably described by
one and the same power-decay curve (Reference curve 2) 
indicating a scale-free behavior for the market dynamics. 
For the Philippine market, the $\Delta
t$-dependence of the $\alpha_{l}$ ($\alpha_{r}$)-values is erratic
with large standard deviations -- there is increasing
asymmetry for $p(\Delta x, \Delta t)$ with $\Delta t$.
Exchange control destroyed the scale-free behavior
of the Philippine market. The largest standard deviation for
Canada and UK is $\pm 0.2$ and $\pm 0.1$, respectively which
happens at $\Delta t = 1$ day, while for Philippines it is $\pm
1.4$ ($\Delta t = 3$ days). For any G7 market, $\langle \Delta x
\rangle$ increases with $\Delta t$ according to a power law since
the exponent is inversely proportional to $1/2\sigma^{2} =
1/\langle \Delta x \rangle$.

The observed power-law dependence of the broadening of $p(\Delta
x, \Delta t)$ with $\Delta t$ is explained as follows. Let
$\langle \Delta x_{Q} \rangle = \langle d_{T} \rangle^{2} $, be
the average fluctuation corresponding to a longer time-delay
$\Delta t_{Q}= Q \Delta t$, where: $d_{T} = | \sum_{q} d_{q}\exp(j
\theta_{q}) |$ $= [( \sum_q d_q \cos\theta_q)^2 + 
(\sum_q d_q \sin\theta_q)^2]^{1/2}$, $\Delta t$ is the basic
time delay, and $q=1,2,\ldots, Q$.  If $\langle d_1 \rangle =
\langle d_2 \rangle = \ldots = \langle d_Q \rangle$, then $\langle
d_T \rangle = \langle d_1 \rangle \langle [( \sum_q\cos\theta_q)^2 
+ ( \sum_q\sin\theta_q)^2]^{1/2} \rangle
\approx Q^{p}\langle d_1 \rangle$.

\begin{figure}
\centering
  \includegraphics[width=\columnwidth]{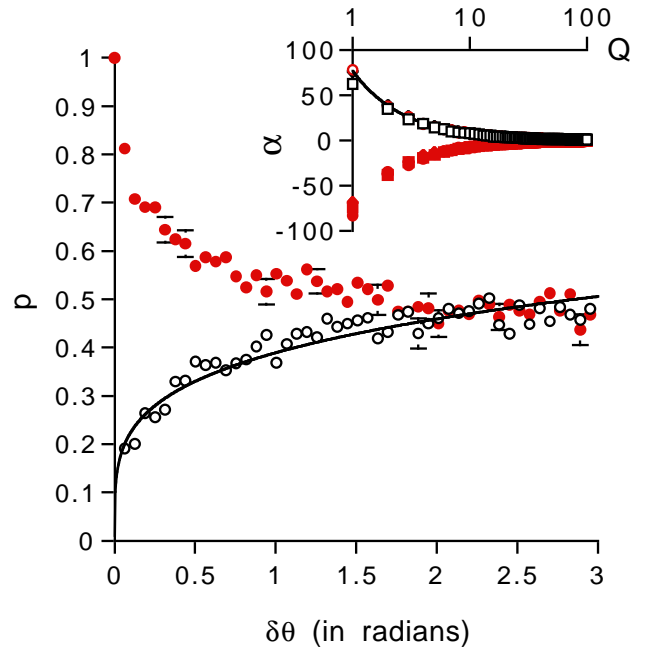}\\
  \caption{
Power $p$ vs. $\delta\theta$. 
Possible directions of $\Theta_q$'s are always forward or always backward (filled circles).
Each datapoint is average of $10$ trials. Reference curve: $p = (\delta \theta)^{0.24}$.
Inset: $\alpha_{l}$ and $\alpha_{r}$ of
$\ln[p(\Delta x_{Q} , Q)]$ vs $Q$ for $Q$ (deg) $= 90$ (squares),
$180$ (circles), and $270$ (diamonds). Reference curve: $\alpha_{l} = 76.6 Q^{-1}$.
  }\label{fig:6}
\end{figure}

The $\langle d_T \rangle$ value depends on the probability
distribution of $\{\theta_q\}$.  If the $\theta_q$'s are
statistically independent and uniformly distributed within
$[0,2\pi]$ then $d_T \approx Q^{1/2}\langle d_1 \rangle$, i.e. $p
\approx 0.5$. Figure~\ref{fig:6} inset plots 
$\alpha_{l}$ and $|\alpha_{r}|$ of $p(\Delta x_{Q}, Q)$
with $Q$ for the above case which verifies our
assumption of a power-law dependence of $\langle d_T \rangle$ with
$\langle d_1 \rangle$.

Figure~\ref{fig:6} plots $p$ as a
function of uncertainty spread $\delta\theta$. 
If $\theta_q$'s are restricted such that
$\theta_{q+1}$ ($q>1$) is uniformly-random within the
forward ($-\pi/2 \leq \theta_q - 0.5\delta\theta \leq
\theta_{q+1} \leq \theta_q + 0.5\delta\theta \leq \pi/2$) and opposite 
($\pi/2 \leq (\theta_q + \pi) - 0.5\delta\theta \leq
\theta_{q+1} \leq (\theta_q + \pi) + 0.5\delta\theta \leq 3\pi/2$) directions
then: $0 \leq p \leq 0.5$.
If the $\theta_q$'s occur only within the forward or opposite direction 
then $0.5 \leq p \leq 1$. In all cases, $\theta_1$ is
uniformly-random in the range $[0,2\pi]$.

A G7 market exhibits a $(\Delta t)^{-0.5}$-dependence (i.e. $p
\approx 1/4$) of $\alpha_l$ (Fig~\ref{fig:5}) because $d_{q+1}$ is
equally likely to be in the same or opposite direction of $d_q$, with 
$\delta\theta \approx 0.3$ (17 deg).  For a market where the
directions of $d_{q+1}$ and $d_q$ are statistically
independent of each other, the decay of $\alpha_l$ with $\Delta
t$ is faster ($p \approx 1/2$, Fig~\ref{fig:6} inset).

In our model, the \forex market consists of $N$ independent
players where $\Delta x$ is treated as a random walk of $N$ steps
in the demarcated complex plane where $|a_n|^2$ represents the
fluctuation $|\Delta x_n|$ of the $n$-th player. Plane anisotropy leads
to asymmetric $p_d(d^2)$'s.
Our model also explains the power-law dependent broadening of the
$p(\Delta x, \Delta t)$'s with $\Delta t$ in G7 markets.

$\Delta x$ is interpreted as the \emph{intensity} of the
\emph{resultant complex amplitude} that arises from the linear
superposition of the fluctuations of the $N$ independent dealers.
Interactions between players arises inherently because the
intensity which is the product of the resultant complex amplitude
and its conjugate, contains interference terms between the
contributions of the individual players.  We showed that real
market dynamics could be analyzed accurately with a relatively low
number of the interacting players. The interaction between the multitude
of agents in a real market could be effectively reduced into
one with a low number of representative classes.

Our model currently neglects the phenomenon of allelomimesis which
causes social agents to cluster\cite{ref18}. 
Herding (bandwagoning) among market dealers could
trigger massive selling or buying in financial markets and leads
to large swings in $\{x(t)\}$ like those found in the middle
1980's for the G7 markets. The same limitation is found in
previous models of \forex markets \cite{ref6,ref7}.

\emph{Acknowledgement.}
J Garcia for stimulating discussions and assistance in
data acquisition (Philippine peso).


\end{document}